# In-plane Dielectric and Magnetoelectric Studies of BiFeO$_3$

**Ashok Kumar[1], J. F. Scott[1,2], R. Martínez[1], G. Srinivasan[3], R. S. Katiyar[1]**

[1]Department of Physics and Institute for Functional Nanomaterials, University of Puerto Rico, San Juan, Puerto Rico, USA, PR-00936-8377

[2]Department of Physics, Cavendish Laboratory, University of Cambridge, Cambridge CB3 OHE, United Kingdom

[3]Physics Department, Oakland University, Rochester, Michigan 48309-4401, USA

In-plane temperature dependent dielectric behavior of BiFeO$_3$ (BFO) as-grown thin films show diffuse but prominent phase transitions near 450 (+/-10) K and 550 K with dielectric loss temperature dependences that suggest skin layer effects. The 450 K anomalies are near the "transition" first reported by Polomska et al. [Phys. Stat. Sol. 23, 567 (1974)]. The 550 K anomalies coincide with the surface phase transition recently reported [Xavi et al. PRL 106, 236101 (2011)]. In addition, anomalies are found at low temperatures: After several experimental cycles the dielectric loss shows a clear relaxor-like phase transition near what was previously suggested to be a spin reorientation transition (SRT) temperature (~ 201 K) for frequencies 1 kHz < f < 1MHz which follow a nonlinear Vogel-Fulcher (V-F) relation; an additional sharp anomaly is observed near ~180 K at frequencies below 1 kHz. As emphasized recently by Cowley et al. [Adv. Phys. 60, 229 (2011)], skin effects are expected for all relaxor ferroelectrics. Using the interdigital electrodes, experimental data and a theoretical model for in-plane longitudinal and transverse direct magnetoelectric (ME) coefficient are presented.

**Corresponding Author:**

E-mail: [#]Ashok Kumar: ashok553@gmail.com,
[*]J F Scott: jfs32@hermes.cam.ac.uk,
R S katiyar: rkatiyar@uprrp.edu

**Introduction:**

Discovery of high polarization, magnetization, piezoelectric coupling, spin wave (magnon) branches, magneto-electric and photovoltaic and exchange bias make BFO one of the widely investigated materials in this decade. [1-16] Most of these properties are studied normal to film plane; but recently Xavi et al. measured the capacitive behavior of BFO single crystal in-plane (surface) by impedance spectroscopy, grazing angle x-ray diffraction (XRD) and x-ray photoelectron spectroscopy (XPS) and found large dielectric anomaly near 550 K. [13] From a device point of view Kumar et al. demonstrated the tuning of spin-wave (magnon) frequencies by 15% for the in-plane configuration under bias electric fields; this opens new possibilities of magnetic logic elements. [15,16] These studies all focus attention upon the interesting properties of surfaces of $BiFeO_3$ and of relaxors. The presence of a skin effect in relaxors was first determined by Noheda et al. in 2001. [17] Further detailed studies on PZN, PMN, and PMN/PT were given by Xu et al. [18,19] and Wong et al.. [20]

Generally these involve thick skin layers, of order 10-50 μm (as in the early study of $BaTiO_3$ by Merz) [21], whereas the present study considers only 150 nm height and 15 μm in-plane distances between electrodes. Important observations from Xu et al. are firstly [18] that the skin layer has the same symmetry but a greater rhombohedral distortion than the bulk (compatible with recent models of iso-structural phase transitions in $BiFeO_3$), and secondly [19] that the skin layer has its own phase transition temperature and displays critical behavior. Xu et al. [22] emphasize that ignorance of these skin layers has led others to erroneous conclusions about the nature of the relaxor state in oxides. The most recent study of skin effects in relaxors is by Wong et al.[20] We note that Xu et al. [19] have shown an unexplained relationship between the direction of polarization (P) in the ferroelectric regions of relaxors and that of P in the adjacent polar nano-regions (PNRs): These polarizations are orthogonal, a fact which was unexplained in. [23] In the present case this could arise from the need to minimize depolarization energy at the surface.

We emphasize that dielectric properties of $BiFeO_3$ films are almost always measured normal to the film plane, but our discovery requires measurement in-plane. The key thing in our work is the use of thin (150 nm) BFO films with pre-patterned interdigitalized (ID) electrodes [procured from NASA Glenn Research Center's electronic division] with 1900 μm (length), 15 μm interdigital spacing, and 150 +/- 25 nm (height) with 45 parallel capacitors in series on sapphire substrates. Scanning electron microscopy (SEM) image of as-grown films is shown in the inset of Fig.1. Growth parameters and electrical measurement set-up are explained in our earlier report. [10, 14] The surface topography across the electrode is shown in the right hand side inset of Fig.1 and indicates average surface roughness ~4-5 nm; this high surface roughness is mainly due to the similar roughness of the substrate (due to electrode processing), and the morphology suggests the fractal grains oriented in different directions. XRD and Raman spectroscopy were used to check the local skin properties across the electrode; different regions across the BFO/ID show defect-free Raman spectra similar to that in single crystal/polycrystalline BFO thin films. [7,9,10] Fig. 1 shows all the 13 first-order Raman-active phonon modes of rhombohedral (R3c space group) crystal structures with 4A1+ 9E modes; in addition to the Raman vibrational modes, one sharp magnon (~ 19.8 $cm^{-1}$ @ 80K) peak was observed up to 460 K (very near the Neel temperature). Magnon softening f(T) will be presented elsewhere. The XRD data matched well with those of polycrystalline BFO thin films, but the bare electrode itself possesses a lot of peaks due to electrode processing. Usually dielectric measurements on ferroelectric thin films have electrodes on the top and bottom surfaces and measure only epsilon normal to the film. Our geometry permitted for the first time accurate measurements of the in-plane dielectric constant versus temperature.

Surface preparation is more important for skin effects. [23] Our observations of temperature and frequency dependences of capacitance and loss are graphically presented in Fig.2 (a)-(f); exceptionally low loss of about ~ 0.1% was observed in all the as-grown samples of the capacitance up to 600 K. Note in these figures that there are two distinct anomalies: One near 450K and the other near 550K. The skin capacitance loss observed in BFO single crystals is very different from that of in-plane thin-film data near the phase transition temperature T*. [13] In addition to the anomalies at high T, the capacitance C(T) showed a broad bump near the lower phase transition temperature (200 K) with very large frequency dispersion in the dielectric loss tangent tan δ (>200 K) for a wide range of frequencies. Apart from this low-temperature feature, a significant dielectric anomaly supported by a loss anomaly was observed at the same temperature 550 (+/- 10) K; and in this case we observed a distinct sharp phase transition is observed for high frequencies (MHz). This is the temperature at which Marti et al. report a surface phase transition.[13] It is worth mentioning that even near the phase transition the dielectric loss remained very low (0.1%). After several experimental cycles the data change and anomalies become smaller. We think that cycling temperature may cause these films to lose some skin effects; either the surfaces or the electrodes may accumulate space charge across the interface and create depolarization fields which in turn give higher capacitance and comparatively high loss (> 500 K), as can be seen in Fig. 2 (c)-(d). We can see the clear dielectric loss dispersion over small temperature regions (ca. 50 K) for wide range of frequencies (1 kHz-1MHz) which follow a nonlinear Vogel-Fulcher (VF) relations (described below). After such repetitive charge injection, the in-plane capacitance eventually loses the high temperature dielectric anomaly (~ 550 K) for high frequencies (>1 kHz). The broad temperature peak only for low-frequency capacitance contradicts the hypothesis of a phase transition confined within the surface or electrode-dielectric [13], suggesting instead a conductivity change, especially since as-grown virgin films without repetitive switching show a well-defined dielectric anomaly. Fig.2 (a-d) data are taken with low oscillation level (OSC) of the impedance analyzer (100 mV) to see the intrinsic effect. Dielectric loss data are considered as a fingerprint in dielectric spectroscopy, and Fig. 2(e)-(f) are taken at a higher OSC level (typically 500 mV), keeping in mind that the distance between the electrodes is 15μm. Note in Figs. 2b,d the presence of an additional anomaly in dielectric loss near the Polomska "transition" at 458K [14]. This anomaly is clear only at low frequencies (kHz) and does not decrease with repetitive cycling. It is studied in more detail elsewhere [16].

In addition to the high-T anomalies near 458K and 540K, our data showed an additional sharp anomaly in the in-plane dielectric loss around 180 K accompanied by high dispersion in dielectric loss. As the data in Figs.2 show, these data would suggest a normal phase transition at 180 K for frequency (< 1kHz) -- perhaps marking the low-T termination of a relaxor phase, a relaxor-like transition between 180 K-235 K (measured at 1 kHz to 1 MHz), and diffuse phase transitions in as-grown films near 458K and 550K (high-frequency values) confined to be in-plane (surface or near-surface). Fig.2 shows that there is a small anomaly in the real part of the in-plane dielectric constant, but this is not very quantitative because of the rapid increase in overall value with rising temperature. The high frequency value agrees well with Raman phonon measurements of Haumont et al. [24] and the very recent grazing-angle XRD studies of Xavi et al. [13], whereas the low-frequency value agrees nicely with that of the previously known transition temperature (200 K) from magnon measurements in thin films by Kumar et al. and in single crystals by Lebeugle et al., & Singh et al. [9, 8, 10].

We were able to do quantitative modeling of the low-temperature phase transition ca.200K (but not to the high-T transitions) to a Vogel-Fulcher model. Our in-plane dielectric loss data after repeated measurements and at different OSC levels showed dielectric dispersion near the previously reported transition temperature near 200K. The temperatures T* at which the dielectric loss exhibited maxima were fitted with a nonlinear VF relation; as shown in Fig. 3, at all OSC levels the imaginary part of the dielectric constant follows a Volger-Fulcher relation for glass-like materials [25, 26, 27]:

$$f = f_0 \exp\left(-E_a / k_B(T^* - T_f)\right)$$ where $f$ is the experimental frequency; $f_0$, the pre-exponential factor; $E_a$, the activation energy; $k_B$, the Boltzmann constant; and $T_f$, the static freezing temperature. This supports the relaxor nature of in-plane BFO thin films. From the non-linear fitting of the imaginary part of dielectric data (Fig. 4) we found $E_a = 0.05(+/-0.01)$ eV, $f_0 = 1 \times 10^8$ Hz, and $T_f = 125$ (+/-10) K for both OSC levels. These parameters are in good agreement with those reported for other relaxors. In particular the value of $T_f$ obtained from Im epsilon can be positive or negative or even non-monatonic, unlike that from Re epsilon, as discussed by Tagantsev [27]. Hence the attempt frequency can differ in the two cases. Its freezing temperature also surprisingly matched with the onset of the spin-glass=like behavior observed in BFO single crystals [28]. Singh et al. showed a gradual enhancement in the zero field cooled (ZFC) and field cooled (FC) magnetization below 140 K with anomalous behavior in ZFC below 100 K; although this was initially suggested to be a spin reordering temperature, spin ordering in other orthoferrites generally involves the rare-earth A-site ion, and this is impossible in bismuth ferrite. [29]

From 2008 onwards, scientists have been searching for the real cause of the 200K transition and onset of spin-glass effects in BFO thin films; a long manuscript has been submitted elsewhere (Dkhil et al., 2012). At present it is uncertain as to whether anomalies at 140.3 K, 201 K, and other below-ambient temperatures are surface or bulk, in-plane or out of plane. The present investigation does not solve this problem, but results indicate sharp dielectric anomaly in the in-plane dielectric loss data near 200K for low frequency probes, relaxor-like behavior for higher frequencies, and a Vogel-Fulcher freezing temperature near the previously determined onset of an apparent BFO spin-glass temperature. This suggests that the phenomena are due to skin effects and in-plane properties of BFO thin films and crystal surfaces. The experiments on magnetic systems have shown a skin effect very similar to that observed for relaxors [23].

The final goal of all these basic or applied research is to get high ME coupling in the single phase BFO. Single-phase or composite materials with piezoelectric and piezomagnetic phase may produce magnetoelectric (ME) signal since $\alpha = \delta P/\delta H$ is the product of both ferroic piezo-(electric/magnetic) properties. There are several ways to measure the ME response: to measure the weak ME signal, a small ac magnetic field $\delta H$ applied to a dc magnetic field bias sample gives a small induced voltage ($\delta V$). The ME voltage coefficient $\alpha_E = \delta V/t\delta H$, and $\alpha = \varepsilon_r \varepsilon_0 \alpha_E$ where t is thickness and $\varepsilon_r$ is the relative permittivity. The paper on BFO by Wang et al. [1] reported the out of plane ME coefficient ($\alpha$) response as high as 3 V/cm·Oe at zero field with further decrease in zero or even negative with increase in dc bias field. In contrast to Wang et al., our in-plane ME responses (both transverse and longitudinal mode) exhibit negligible ME response without a dc bias field, with further increase in ME response with increase in dc bias. The so-called characteristic peak in response (versus H) was also observed at 1.4 kOe dc with unsaturated peak in longitudinal ME data within the experimental limitation [30, 31]. A surprisingly similar ME response was observed in PZT/ $Co_{1-x}Zn_xFe_2O_3$ (CZFO) hetrostructures in which a flat ME peak was observed compared to the transverse ME response. We believe the main reason is the suppression of the spin spiral in nanoscale as observed in the $BiFeO_3$ nanocrystals and thin films [10,32,33]. Our films also showed only one magnon (spin waves) despite the existence of several electromagnons observed in BFO single crystals [8,9,15]. Suppression of the spin cycloid at nanoscale [32] yields large magnetization(~ 0.4 $\mu_B$/Fe), which in turn gives rise to an enhanced non-ferroelectric rotation of oxygen octahedra about the [111] axis via Dzyaloshinskii-Moriya interaction [32]. Goswami et al. showed large off-center distortion ($\delta \sim 0.06$ Å) which is equivalent to suppression of polarization by 7% in the $Bi^{+3}$ ions of BFO nanocrystals under magnetic field (5 T) that produces a large ME coupling at nanoscale [33]. Ederer et al. emphasized that the polar and rotational distortions are coupled in BFO at nanoscale which significantly influences the polarization and hence the ME coupling [34]. One more factor that may influence the ME signal is coupling of ferroelectric domains and magnetic domains at nanoscale via (piezo)striction, as seen in hexagonal manganites [35].

**Theoretical Modeling for Single-phase ME multiferroics**: Most of the model for ME coupling is developed for layered composites without any substrate clamping [30,31,36]. In-plane BFO may be considered as a free body and its ME response should follow the same behavior as composites. BFO possesses a reasonably good piezoelectric coefficient ($d_{33}\sim$ 50pV/m),[37] strain (piezo/magnetic ~ $10^{-6}$), [38] photostrictions (10-14 ppm), but of course its values change significantly in thin films. We first started simulation with primary data posted in the literature for piezostriction and magnetostriction [31-39]: the fitted data are; $^p s_{xx} = {^p}s_{xy} = {^m}s_{xx} = {^m}s_{xy} \sim 10^{-11}$ $m^2$/N, $\varepsilon_{33}^{T,p} \sim$ 2 x $10^{-9}$ F/m, n=3.4 for transverse mode and $^p s_{xx} = {^p}s_{xy} = {^m}s_{xy} = {^m}s_{xy} \sim 10^{-11}$ $m^2$/N, $\varepsilon_{33}^{T,p} \sim$ 2 x $10^{-9}$ F/m, n=1.5 for longitudinal mode. All the piezoelectric and magnetostrictive compliance factors are kept identical for the BFO single phase system. The numerical values of the data are meaningful and matched with some of the experimentally reported values, although we found large discrepancies in the numerical values of these physical parameters in the literature due to different strain state and crystal structure. BFO crystal showed tensile strain [38] in transverse mode, as observed in the present case, and compressive in longitudinal ME mode. We have assumed that no electric field is present in the magnetostrictive (*m*) layer, an equipotential piezoelectric layer, uniform magnetic field, and no in-plane clamping for the model.

At the atomic scale, the Harshe model can be used for single-phase compounds having strong electromechanical and megnetomechanical interactions. ME coefficients ($\alpha_E$) for transverse (in-plane mode) and longitudinal (out-of-plane mode) are represented respectively as:[28].

$$\alpha_{E,31} = \partial E_3 / \partial H_1$$
$$= \frac{-2k\nu(1-\nu)d_{31}^p q_{11}^m}{(s_{11}^m + s_{12}^m)k\varepsilon_{33}^{T,p}\nu + (s_{11}^p + s_{12}^p)\varepsilon_{33}^{T,p}(1-\nu) - 2k(d_{31}^p)^2(1-\nu)} \quad (1)$$

And

$$\alpha_{E,33} = \partial E_3 / \partial H_3$$
$$= \frac{-2\nu(1-\nu)d_{13}^p q_{31}^m}{(s_{11}^m + s_{12}^m)\varepsilon_{33}^{T,p}\nu_p + (s_{11}^p + s_{12}^p)\varepsilon_{33}^{T,p}(1-\nu) - 2(d_{31}^p)^2(1-\nu)} \quad (2)$$

Here p and m represent parameters for piezoelectric and magnetostrictive phases; $s$ and $d$ are compliance and piezoelectric coefficients; $\varepsilon^T$ is the permittivity at constant stress; $q$, piezomagnetic coefficients; and $\nu = {}^p\nu/({}^p\nu + {}^m\nu)$, $\nu^p$ and $\nu^m$ denote the volume of piezoelectric and magnetostrictive phases (we consider similar volumes for both the ferroic phases), respectively. Magnetostrictive strain induced in the piezomagnetic phase should be fully transferred to the piezoelectric phase. Equations (1) and (2) describe the dependence of magnetoelectric voltage coefficient on piezomagnetic coefficient $q$. The piezomagnetic coefficient $q$ is obtained from $\delta\lambda/\delta H$, where $\lambda$ is the magnetostriction. Therefore, any dependence of the magnetoelectric coefficients on field orientation results from the magnetostriction. Our model used a sigmoidal Hill function for the magnetostriction. Thus the relationship between static magnetostriction and applied field can be given by:

$$\lambda = A \cdot \frac{H^n}{K^n + H^n} \quad (3)$$

Here A, K and n are adjustable parameters.
The magnetostriction and piezomagnetic coefficients are obtained from the fitted parameters matched with the earlier experimental values for very low bias magnetic field [39]. BFO thin films possess comparatively low magnetostriction values (~ $10^{-14}$ to $10^{-10}$ depending upon the probe dc bias magnetic field) compare to their bulk counterpart.

**Summary:**

Four novel in-plane phase transitions are established: The first is at 180 K and observed at very low frequencies (< 1 kHz) with a sharp dielectric loss anomaly; the second is a relaxor-like phase transition (180-235K); the third is a diffuse phase transition at 550 K. The fourth is the Polomska transition at 458K. The first two may represent a single Vogel-Fulcher behavior with onset upon cooling at 230-5 K, depending upon probe frequency, and completion at 180K. The above-ambient transition near 450K coincides with the Polomska anomaly and like the better studied skin-transition near 540K is probably a surface transition. The new work with planar electrodes also permits measurement of reasonable good ME coefficients (~ 300 mV/Oe.cm in transverse and 2 V/Oe.cm in longitudinal). Understanding of the in-plane phase transitions and ME behaviour may be helpful to design nanoscale and interfacial devices with high surface to volume ratios.

**Acknowledgement:**

This work was partially supported by EPSRC (at Cambridge), W911NF-06-1-0183, W911NF1110204 and DoE FG 02-08ER46526 grants.

**References:**

[1] J. Wang, J. B. Neaton, H. Zheng, V. Nagarajan, S. B. Ogale, B. Liu, D. Viehland, V. Vaithyanathan, D. G. Schlom, U. V. Waghmare, N. A. Spaldin, K. M. Rabe, M. Wuttig and R. Ramesh, *Science* **2003**, 299 , 1719.

[2] R. J. Zeches, M. D. Rossell, J. X. Zhang, A. J. Hatt, Q. He, C.-H. Yang, A. Kumar, C. H. Wang, A. Melville, C. Adamo, G. Sheng,Y.-H. Chu, J. F. Ihlefeld, R. Erni, C. Ederer,V. Gopalan, L. Q. Chen5, D. G. Schlom, N. A. Spaldin, L. W. Martin and R. Ramesh, *Science* **2009**, 326, 977.

[3] T. Choi, S. Lee, Y. J. Choi, V. Kiryukhin, S.-W. Cheong, *Science*, **2009**, 324 (3), 63.

[4] W. Eerenstein, N. D. Mathur & J. F. Scott, *Nature (London)* **2006**, 442, 759.

[5] G. Catalan and J. F. Scott, *Adv. Mater.* **2009**, 21, 2463.

[6] D. Lebeugle, A. Mougin, M. Viret, D. Colson, and L. Ranno, *Phys. Rev. Lett.* **2009**, 103, 257601.

[7] R. Palai, R. S. Katiyar, H. Schmid, P. Tissot, S. J. Clark, J. Robertson, S. A. T. Redfern, G. Catalan, and J. F. Scott *Phys. Rev. B* **2008**, 77, 014110.

[8] D. Lebeugle, D. Colson, A. Forget, M. Viret, A. M. Bataille, and A. Gukasov, *Phys. Rev. Lett.* **2008**, 100, 227602.

[9] M. K. Singh, Ram S Katiyar, J F Scott, *J. Phys. Condens. Matter* **2008**, 20, 252203.

[10] A. Kumar, N. M. Murari, and R. S. Katiyar, *Appl. Phys. Lett*. **2008**, 92, 152907.

[11] S. M. Wu, Shane A. Cybart, P. Yu, M. D. Rossell, J. X. Zhang, R. Ramesh, R. C. Dynes, *Nature Mater.* **2010**, 9, 756.

[12] F. Kubel and H. Schmid, *J. Cryst. Growth* **1993**, 129, 515.

[13] X. Marti, Pilar Ferrer, Julia Herrero-Albillos, Jackeline Narvaez, Vaclav Holy, Nick Barrett6, Marin Alexe, and Gustau Catalan, *Phys. Rev. Lett*, **2011**, 106, 236101.

[14] M Polomska, W. Kaczmare, Z. Pajak, *Physica Status Solidi A* **1974**, 23, 567-574.

[15] A. Kumar, J. F. Scott, R. S. Katiyar, *Appl. Phys. Lett.* **2011**, 99, 062504.

[16] A. Kumar, J. F. Scott, R. S. Katiyar, ArXiv:1202.1040

[17] B. Noheda, D. E. Cox, G. Shirane, S E Park,, L E Cross, and Z Zhong, *Phys Rev Lett* **2001**, 17, 3841.

[18] G. Xu, H. Hiraka, G. Shirane, K Ohwada, *Appl. Phys. Lett.* **2004**, 84, 3975.

[19] G. Xu, P M Gehring, C . Stock, K. Conlon, *Phase Trans.* **2006**, 79, 135.

[20] K. S. Wong, B. Wang, J-Y Dai, and H Luo, *IEEE Trans. Ultrson. Ferro. and Freq. Cntrl.* **2008**, 55, 952.

[21] W. J. Merz, *Phys. Rev*. **1954**, 95, 690.

[22] G. Xu, P. M. Gehring, and G. Shirane, *Phys. Rev. B* **2006**, 74, 104110.

[23] R. A. Cowley, S. N. Gvasaliyac, S. G. Lushnikovd, B. Roesslic & G. M. Rotaruc., *Advances in Physics,* **2011**, 60, 229.

[24] R. Haumont, J. Kreisel, P. Bouvier, and F. Hippert, *Phys. Rev. B* **2006**, 73, 132101.

[25] A. A. Bokov and Z.-G. Ye, *J. Mater. Sci.*, **2006**, 41, 31.

[26] H. Vogel, *Z. Phys.* **1921**, 22, 645.

[27] A. K. Tagantsev, *Phys. Rev. Lett.* **1994**, 72, 1100.

[28] Manoj K. Singh, W. Prellier, M. P. Singh, Ram S. Katiyar, and J. F. Scott, Phys. Rev. B *2008*, 77, 144403.

[29] Yu. G. Chukalkin and B. N. Goshchitskii, *Phys. Status Solidi A* **2003**, 200, R9.
[30] J Ryu, S Priya, K Uchino and H E Kim *J. Electroceram.* **2002**, 8 107.
[31] M. I. Bichurin, V. M. Petrov, and G. Srinivasan, Phys. Rev. B **2003**, 68, 054402.
[32] R. Mazumder, P. S. Devi, D. Bhattacharya, P. Choudhury, A. Sen, and M. Raja, Appl. Phys. Lett. **2007**, 91, 062510.
[33] Sudipta Goswami, Dipten Bhattacharya, P. Choudhury, B. Ouladdiaf, and T. Chatterji *Appl. Phys. Lett.* **2011**, 99, 073106.
[34] C. Ederer and N. A. Spaldin, *Phys. Rev. B* **2005**, 71, 060401(R).
[35] T.-J. Park, G. C. Papaefthymiou, A. J. Viescas, A. R. Moodenbaugh, and S. S. Wong, *Nano Lett.* **2007**, 7, 766.
[36] B. Ruette, S. Zvyagin, A. P. Pyatakov, A. Bush, J. F. Li, V. I. Belotelov, A. K. Zvezdin, and D. Viehland, *Phys. Rev. B* **2004**, 69, 064114.
[37] R. J. Zeches, M. D. Rossell, J. X. Zhang, A. J. Hatt, Q. He, C.-H. Yang, A. Kumar, C. H. Wang, A. Melville, C. Adamo, G. Sheng, Y.-H. Chu, J. F. Ihlefeld, R. Erni, C. Ederer, V. Gopalan, L. Q. Chen, D. G. Schlom, N. A. Spaldin, L. W. Martin, R. Ramesh, *Science* **2009**, 326, 977.
[38] B. Kundys M. Viret, D. Colson & D. O. Kundys, *Nature Materials*, **2010,** 9, 803.
[39] A. K. Zvezdin, A. M. Kadomtseva, S. S. Krotov, A. P. Pyatakov, Yu. F. Popov, G. P. Vorob'ev, *Journal of Magnetism and Magnetic Materials* **2006**, 300, 224.

**Figure captions:**

Fig.1 Micro-Raman spectra of BFO thin films at 80 K, showing all 13 first order Raman-active phonon modes with well defined magnon mode at 19.9 cm$^{-1}$; inset left hand side is the SEM picture of the device used to measure the in-plane dielectric and ME response; bar line indicates the 500 μm length; right inset is the surface topography of the BFO thin film with polygranular structure; the scale of the topography lies from 0-100 nm, and the morphology of the grains are fractal in nature with average length 40-60 nm.

Fig.2 In-plane dielectric response of the 100-nm-thick BFO thin films grown on pre-patterned interdigital electrodes: (a) capacitance behavior of as-grown films show prominent diffuse dielectric peaks near 550 K, and the dielectric loss data also support this finding; (b) tan δ indicates a wide dielectric temperature dispersion as a function of frequency (< 100 K below 100 kHz) and a prominent peak near the capacitive anomaly. Note also a peak near 450K (Polomska anomaly) at 1 kHz; (c) In-plane capacitance measurements on the same sample illustrate dielectric anomaly near 550 K for low probe frequency (inset 50 kHz-1kHz) and very high dielectric dispersion for higher frequency. (d) 50 K temperature dispersion as a function of frequency in tan δ can be seen above 1 kHz, whereas a sharp loss anomaly at 180 K is observed at the lower frequencies (for 50 Hz to 1kHz). Note that an anomaly exists near 450 K also; (a)-(d) represent data at 100 mV OSC level on the impedance analyzer. (e) Sharp anomaly in dielectric loss at 180 K and loss temperature dispersion as a function of frequency of about 50 K for higher frequencies at 500 mV OSC.

Fig.3 Nonlinear VF equation fitting (solid line) of the experimental data (dotted line) obtained from dielectric loss data for the transition near 200K; the peak position of the loss data are obtained with a least squares fit. The relaxation time obtained in both cases is approximately 10$^{-9}$ s which physically matches typical relaxor-like systems. The freezing temperature is 125 (+/-5)K and represents the onset of a spin-glass phase in BFO.

Fig. 4 The in-plane transverse and longitudinal ME coefficient behavior as functions of dc bias field; an ac magnetic field of 1 Oe at 1 kHz was used to generate the ME voltage. The transverse ME response shows maxima near 1400 Oe similar to other piezoelectric/piezomagnetic hetrostructures (i.e. PZT/CZFO), whereas the longitudinal ME response displays unsaturated behaviour under dc magnetic bias field.

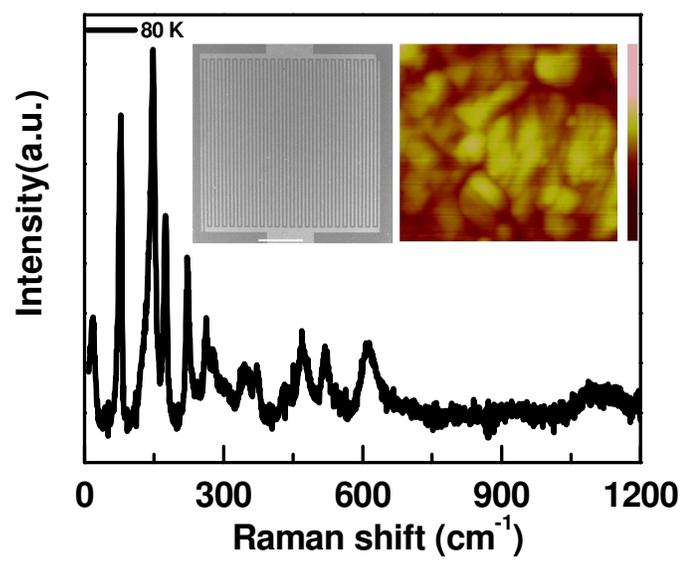

**Fig.1**

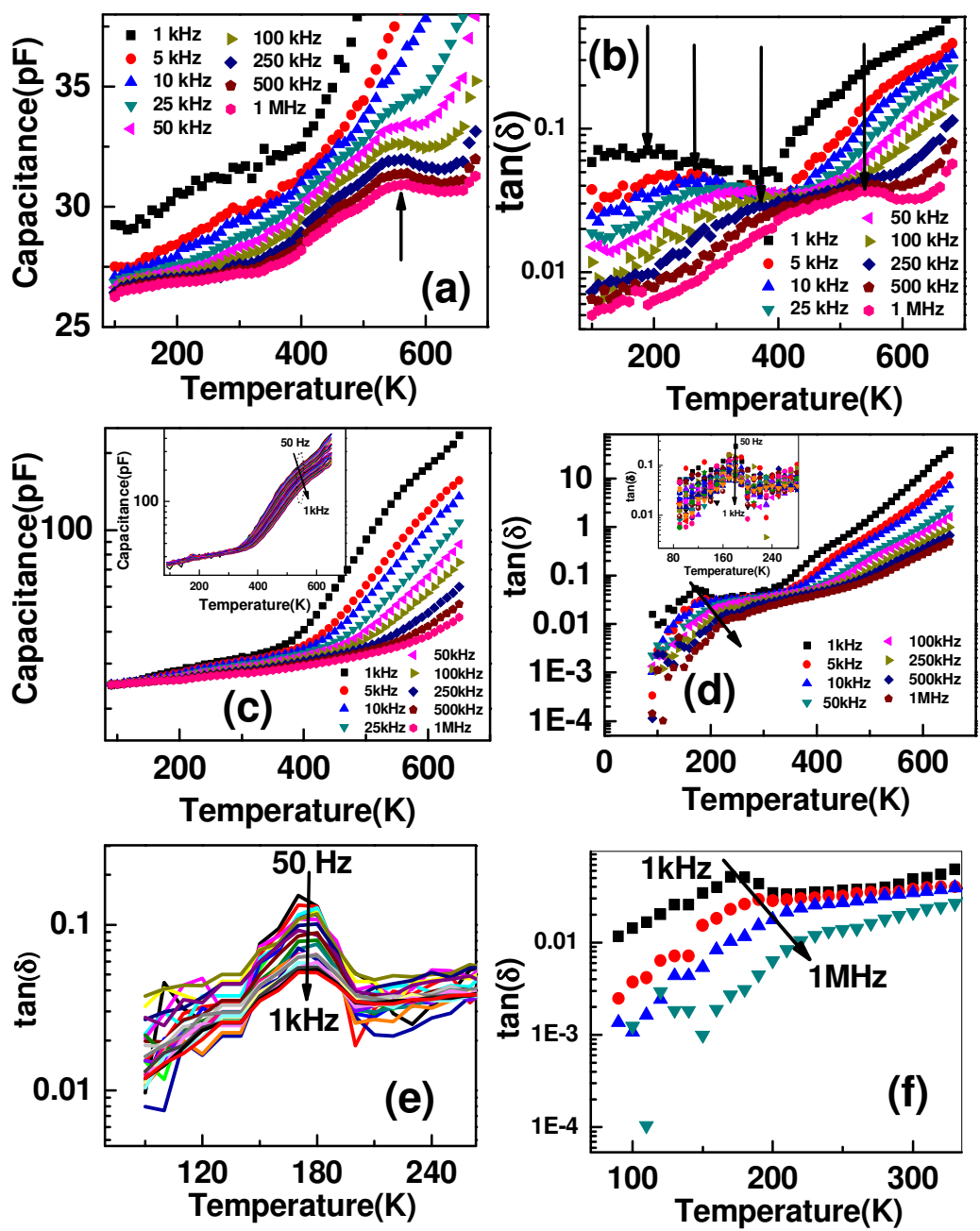

Fig.2

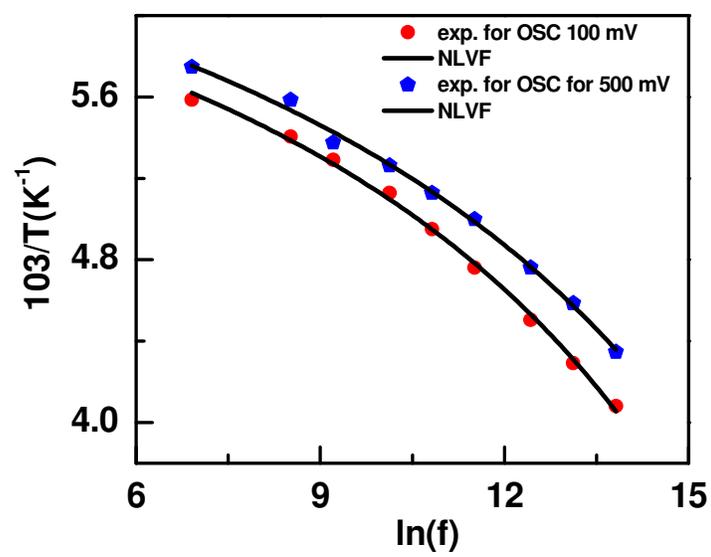

**Fig.3**

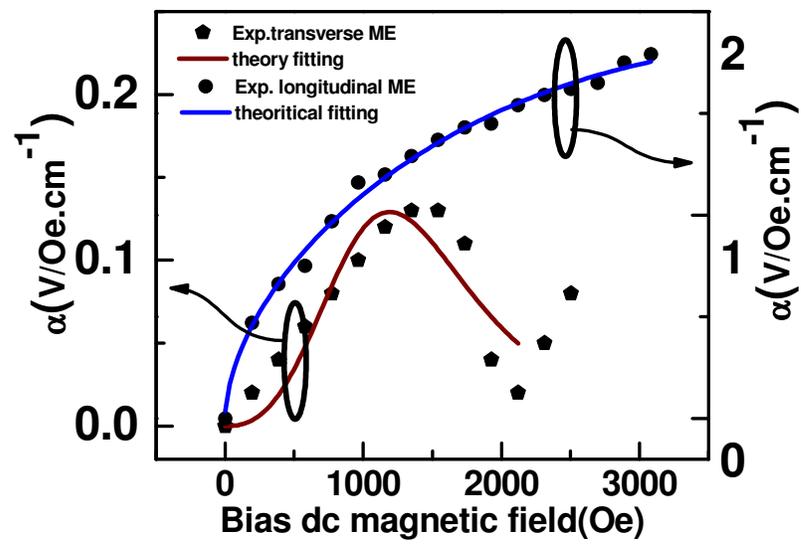

**Fig.4**